\documentclass[iop]{emulateapj}%{mn2e}
\usepackage{amssymb}
\usepackage{natbib}
\usepackage{graphicx}
\def\kms{{\rm \thinspace km \thinspace s}^{-1}}

\def\Msun{\hbox{$\rm\thinspace M_{\odot}$}}
\def\pc{{\rm\thinspace pc}}     
\def\s{{\rm\thinspace s}}

\def\Myr{{\rm\thinspace Myr}}

\def\K{{\rm\thinspace K}}

\def\ndens{\thinspace\mathrm{cm}^{-3}}

\def\mnras{MNRAS}
\def\apj{ApJ}
\def\aap{A\&A}
\def\apjl{ApJL}
\def\apjs{ApJS}

\shorttitle{Oscillating Filaments: I}
\shortauthors{Matthias Gritschneder, Stefan Heigl, Andreas Burkert}

\begin{document}

\title{Oscillating Filaments: I - Oscillation and Geometrical Fragmentation}

\author{Matthias Gritschneder$^{1}$, Stefan Heigl$^{1}$, Andreas Burkert$^{1,2}$}
\affil{$^1$ University Observatory Munich, LMU Munich, Scheinerstrasse 1, 81679 Munich,
Germany \\
$^2$ Max-Planck-Fellow, Max-Planck-Institute for Extraterrestrial
Physics, Giessenbachstrasse 1, 85758 Garching, Germany} 
\email{gritschm@usm.uni-muenchen.de}

\begin{abstract}
We study the stability of filaments in equilibrium between gravity and internal as well as external pressure using the grid based AMR-code RAMSES. A homogeneous, straight cylinder below a critical line mass is marginally stable. However, if the cylinder is bent, e.g. with a slight sinusoidal perturbation, an otherwise stable configuration starts to oscillate, is triggered into fragmentation and collapses. This previously unstudied behavior allows a filament to fragment at any given scale, as long as it has slight bends. We call this process `geometrical fragmentation'. In our realization the spacing between the cores matches the wavelength of the sinusoidal perturbation, whereas up to now, filaments were thought to be only fragmenting on the characteristical scale set by the mass-to-line ratio. Using first principles, we derive the oscillation period as well as the collapse timescale analytically. To enable a direct comparison with observations, we study the line-of-sight velocity for different inclinations. We show that the overall oscillation pattern can hide the infall signature of cores.
\end{abstract} 

\keywords{ISM: kinematics and dynamics, ISM: structure, stars: formation, hydrodynamics, methods: analytical, methods: numerics}

\section{Introduction}
Observations with the Herschel satellite have revealed that the backbone of molecular clouds is a complex network of connecting and interacting filaments \citep[e.g.][]{Andre:2010,Molinari:2010,Arzoumanian:2011}. The dynamics of this web of dense  molecular gas not only determines the evolution and stability of molecular clouds, but also regulates their condensation into stars. Molecular cloud cores and single low-mass stars are almost always found in filaments, often aligned like pearls on a string \citep{Hartmann:2002,Lada:2008,Andre:2010}. 
Molecular clouds themselves are part of a larger web-like network of gas structures
on galactic scales with high-density regions, embedded in diffuse and hot atomic and ionized
inter-clump material that determines the average gas pressure of the ISM \citep{Ostriker:2010}.
\begin{figure*}
\begin{center}
{\centering 
\includegraphics[width=21cm]{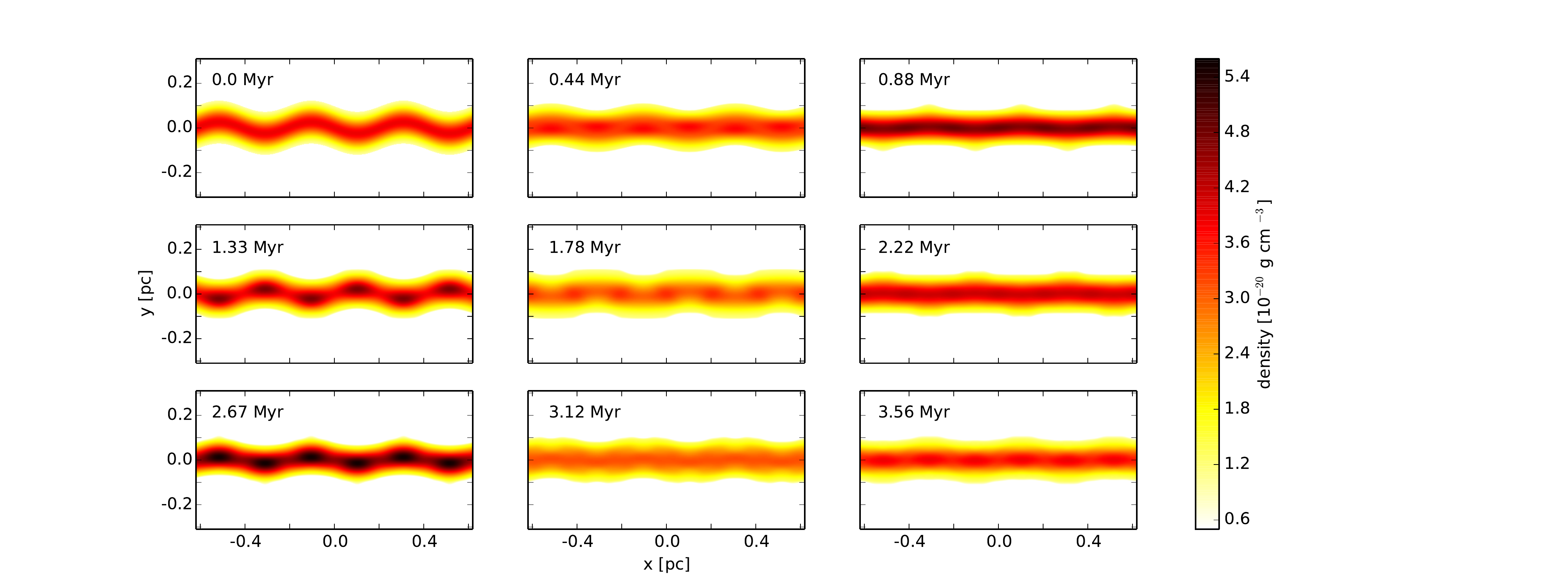}
}
\end{center}
\caption{Time evolution of our fiducial simulation. Due to its own gravity a bent filament initially at rest starts to oscillate. The time increases from top left to bottom right.  Our fiducial simulation has 3 periods of the sinusoidal perturbation and $f_{\rm cyl}=0.5$.  \label{fig:sims_osc}}
\end{figure*}

Locally, progress has been made by simulating colliding flows as a method of cloud or filament formation.
E.g. \citet{Heitsch:2006} studied the formation of cold and dense clumps between two infinite
flows which collide on a perturbed interface. 
Their study showed that cold structure can arise even from initially uniform flows if the
conditions favor certain fluid instabilities, such as the Non-linear Thin Shell Instability (NTSI, \citealt{Vishniac:1994}),
the Thermal Instability \citep{Field:1965} and the Kelvin-Helmholtz instability. 
\citet{Vazquez-Semadeni:2007} used colliding cylindrical flows with lengths of tens of parsecs, 
adding random velocity perturbations to the average flow velocity.  
This allowed them to study star formation efficiencies and the fates of individual clouds.
More recently, it was shown that the newly found, complex filaments \citep{Hacar:2013} can be readily explained by the evolution of the turbulent ISM \citep{Moeckel:2015aa}. In addition, studies on the width of filaments have been performed \citep{Smith:2014aa,Federrath:2015aa,Seifried:2015aa}. The evolution of infinite, straight filaments has been investigated in great details. For example, \citet{Heigl:2016} showed that the observed densities and velocities in L1517 \citep{Hacar:2011} can be readily explained by the evolution of an inclined infinite filament initially in hydrostatic equilibrium. The gravitational evolution of filaments with finite length has also recently been explored \citep[e.g.][]{Pon:2012}. \citet{Keto:2014} showed that finite filaments collapse preferentially parallel to their long axis leading to oscillating cores which can explain the complex velocity pattern seen in observations. \citet{Clarke:2015}, following \citet{Burkert:2004}, determined the longitudinal collapse timescale of finite filaments and their end-dominated fragmentation.

Here, we extend the previous studies parametrizing a filament as a cylinder in hydrostatic equilibrium \citep[e.g.][]{Stodolkiewicz:1963aa,Ostriker:1964aa} by adding a geometrical bend or perturbation to it. In this first paper we focus on infinite cylinders. A discussion of the evolution of finite, curved filaments is  postponed to a subsequent paper of this series. First, we review the basic equations in \S \ref{theory}. In \S \ref{sims}, we show that these filaments start to oscillate, derive the oscillation frequency and show that above a given initial density this filaments or cylinders fragment. To enable a direct comparison to observations, we discuss the line-of-sight velocities of our simulations. We summarize in \S \ref{summary}.
\section{Basic equations}
\label{theory}
Infinite cylinders have long been a topic of interest in the context of star formation \citep{Stodolkiewicz:1963aa,Ostriker:1964aa}. Here, we follow the approach by  \cite{Fischera:2012aa} and also \cite{Heitsch:2013a,Heitsch:2013b}. A filament is prescribed by a infinitely long self-gravitating cylinder in thermal equilibrium. The density profile is then given by
\begin{equation}
\rho(r) = \frac{\rho_c}{(1+(r/\sqrt{8}r_0)^2)^2}
\end{equation}
with $r_0=c_s^2/(4\pi G\rho_c)$, where $c_s$ is the isothermal sound-speed, $G$ is the gravitational constant and $\rho_c$ is the central density. Another important quantity is the line mass. It can be shown that the line mass asymptotically reaches a maximum value of 
\begin{equation}
\left(\frac{M}{L}\right)_{\rm crit} = \frac{2c_s^2}{G} = 16.4 \left(\frac{T}{10\K}\right)\Msun\pc^{-1}.
\end{equation}
This maximum value is the critical value, corresponding to an infinite central pressure. A filament with this value for the line mass would collapse immediately, as there is no stable solution. We use this critical mass to define the parameter $f_{\rm cyl}$ via
\begin{equation}
\left(\frac{M}{L}\right) = f_{\rm cyl}\left(\frac{M}{L}\right)_{\rm crit}.
\end{equation}
Below, we perform simulations with a different values of $f_{\rm cyl}$. This is achieved by keeping the central density constant and cutting the profile at a value $\rho_b$ corresponding to the desired $f_{\rm cyl}$. Outside of this radius, we continue with a low density ambient medium in pressure equilibrium with the boundary. 

\section{Simulations}
\label{sims}
In order to test the stability of a filament and to study potential core formation we chose a novel approach. Instead of perturbing the density we apply a geometrical perturbation, e.g. a sinusoidal perturbation in the y direction and evolve the setup with the AMR-grid code RAMSES  \citep{Teyssier:2002}.
\subsection{Initial conditions} 
For all simulations we set up a filament in hydrostatic equilibrium embedded in a warm, ambient medium. We then add the geometrical perturbation in the y direction. Our fiducial case has a central number density of $n_c = 1\times 10^4\ndens$ with $\mu=2.36$. We chose $f_{\rm cyl}=0.5$ and the simulation domain is ${\it l} = 1.23\pc$. This length corresponds to two times the length scale of the fastest growing mode for the case of a perturbation in density. However, here we do not perturb the density, but the geometry. The simulations are performed with a minimum level of 5 and a maximum refinement level of 9, the `hllc' solver and a Courant factor of $0.8$. A new level of refinement is started whenever the error in the gradient of the density becomes larger than $5\times 10^{-2}$.
We add a sinusoidal perturbation in the y direction with three periods, the amplitude is $2\%$ of the box length. The temperature is assumed to be constant with a fixed value in the cold and in the warm, ambient gas. The equation of state is calculated with $\gamma\approx1$ The boundary conditions are periodic in all directions. 
 In  Figure \ref{fig:sims_osc}, we show the result of the simulations with $f_{\rm cyl}=0.5$.  It can be clearly seen that the filament starts to oscillate. This is due to the gravity of the rest of the filament pulling the maxima and minima of the sine towards the mid-plane. As the momentum carried by the ambient medium is small, the system is only slightly damped and enters a stable oscillation.
\subsection{Oscillation and Damping}
\begin{figure}
\begin{center}
{\centering 
\includegraphics[width=7cm]{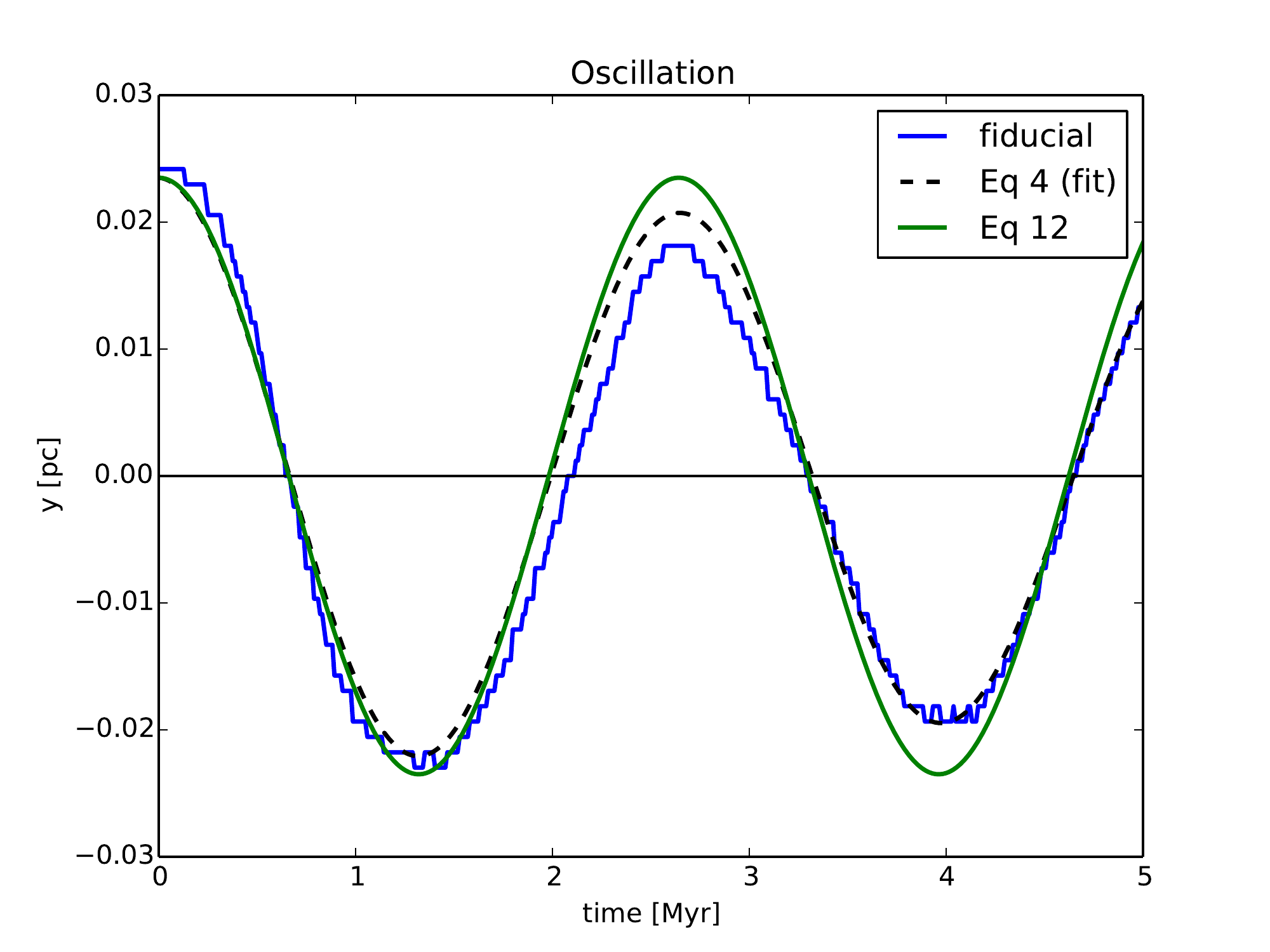}
\includegraphics[width=7cm]{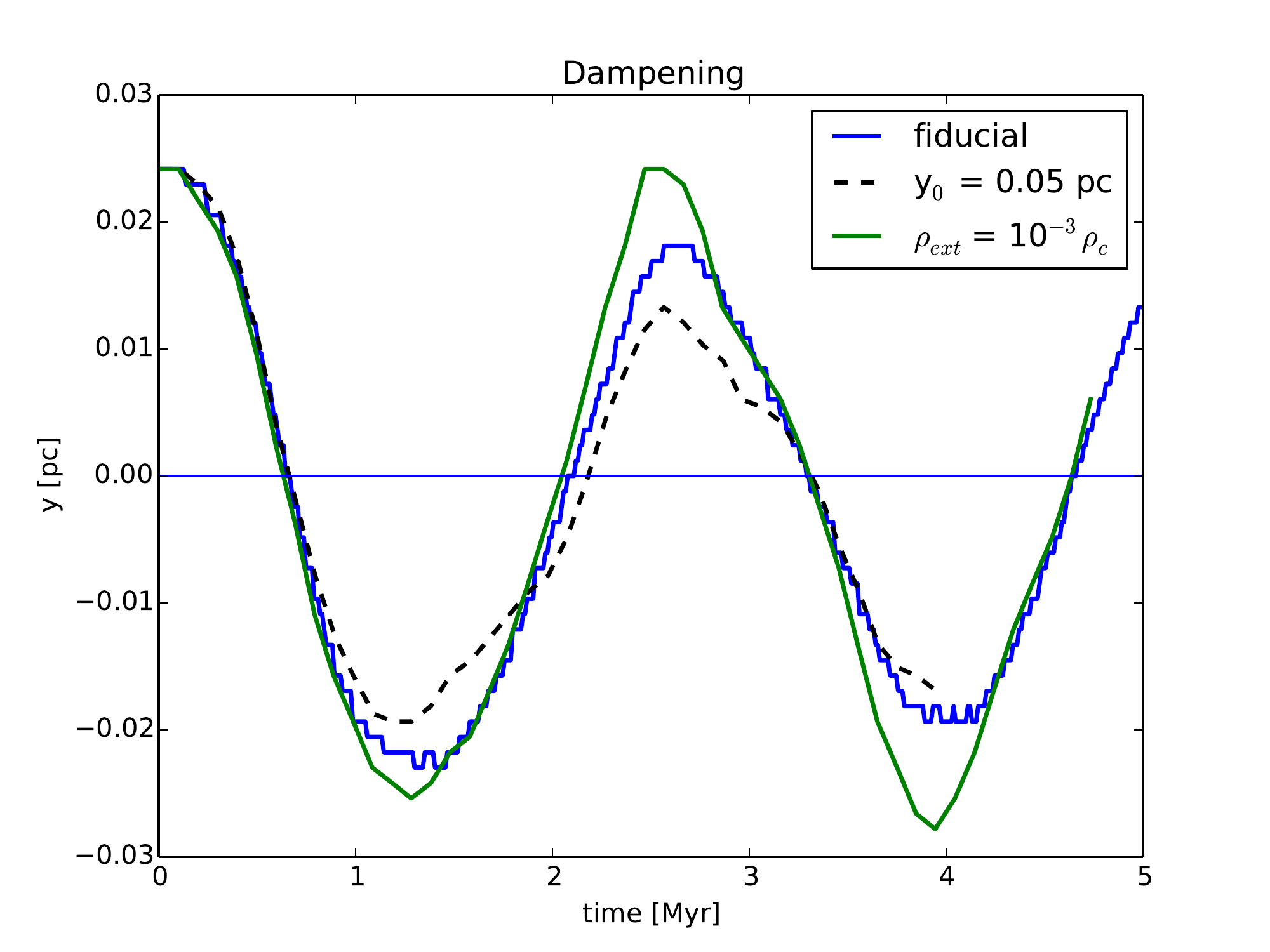}
}
\end{center}
\caption{Position vs time for a peak in the y perturbation. Top: The oscillation of our fiducial filament (blue). Over-plotted are the fit from Eq 4 (black dashed) and the derived solution according to Eq 12 (green). Bottom: The damping effect of the surrounding medium. The fiducial simulation (blue) shows less damping than the same case with double the initial amplitude (rescaled as black dotted line). The case with a ten times lower density in the surrounding medium (green) is virtually un-damped. \label{fig:sims_evol}}
\end{figure}

We show the oscillation in more detail in Figure \ref{fig:sims_evol}. To estimate the position of the filament at a given time, we take the profile at the position of the first upwards peak in y and assume the filament is the region with a density higher than $\rho_c/5$. We then plot the mean value of y as a solid blue line in  Figure \ref{fig:sims_evol} (top panel). It directly shows that the filament is indeed undergoing a damped oscillation.
\begin{figure}
\begin{center}
{\centering 
\includegraphics[width=7cm]{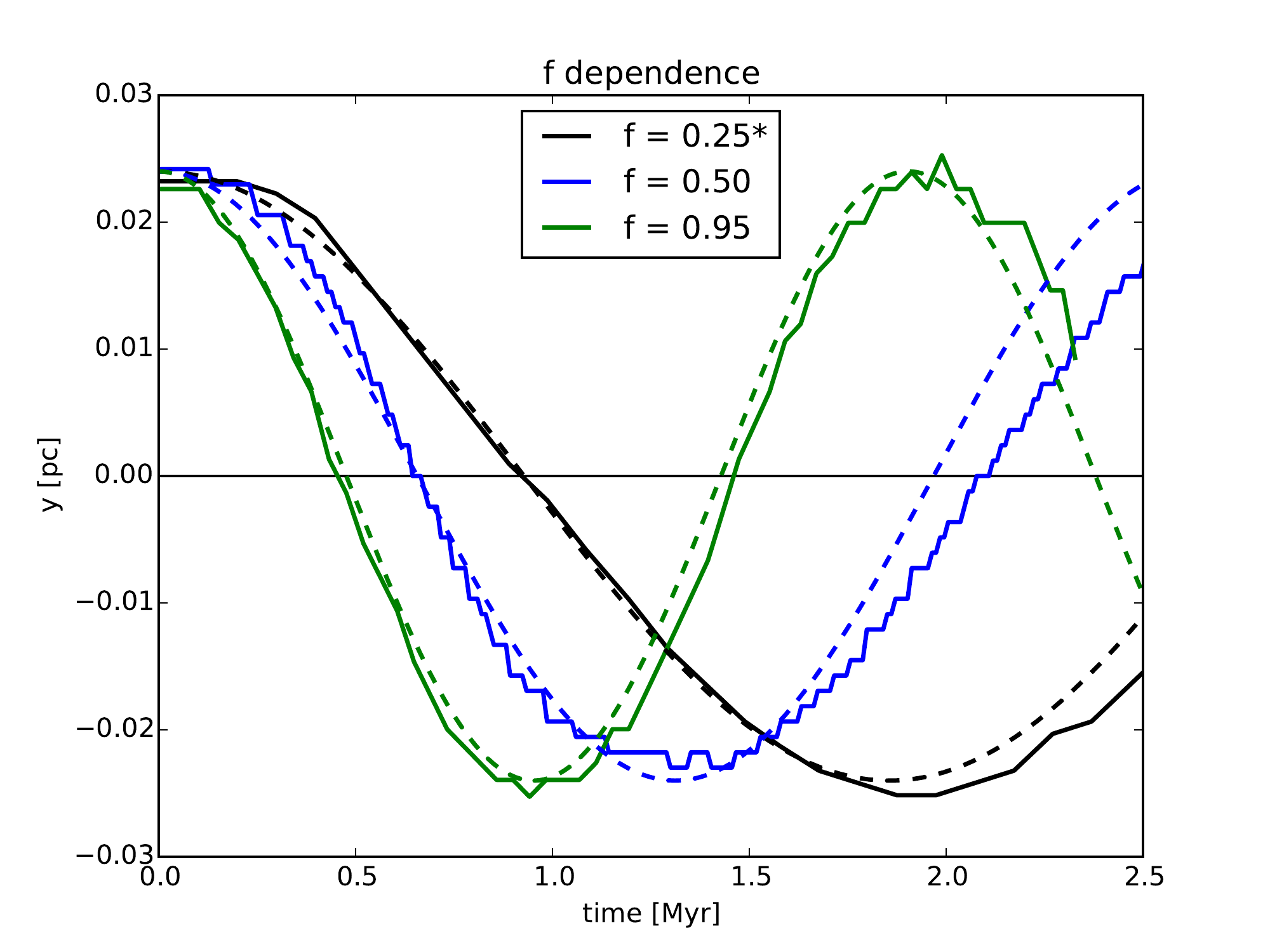}
\includegraphics[width=7cm]{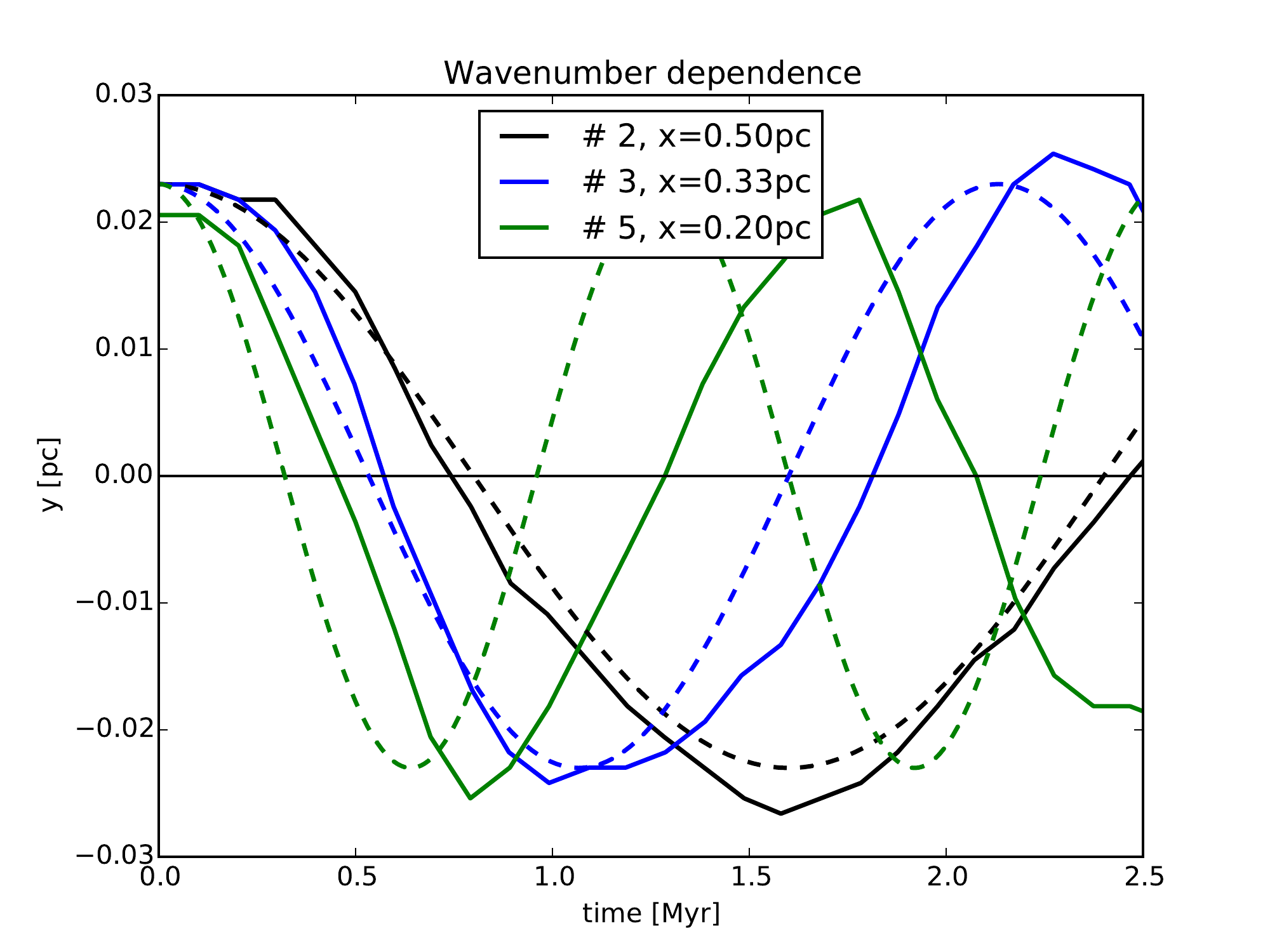}
}
\end{center}
\caption{ Position vs time of a peak in the parameter study. Top: The dependence on $f_{\rm cyl}$. The simulations are plotted as solid lines, the derived solutions (Eq. 12) as dashed lines. The agreement is encouraging. Bottom: The dependence on the wavenumber ${\it W}$ (and thereby the period)  of the sinusoidal perturbation. Again, the simulations are the solid lines, the derived solutions (Eq. 12) are the dashed lines. In this case the agreement is good for lower wave-numbers ${\it W}$. Large deviations are seen for higher wave-numbers. \label{fig:sims_ps}}
\end{figure}
We prescribe this oscillation by
\begin{equation}
y(t) = y_0\thinspace e^{-\delta t}cos(\sqrt{(\omega^2-\delta^2)}t).\label{eq_fit}
\end{equation}
A reasonable fit for $\omega = 7.5\times 10^{-14}\s^{-1}$ and $\delta = 1.5\times 10^{-15}\s^{-1}$ is shown as black dashed line in  Figure \ref{fig:sims_evol} (top panel). While it is encouraging that the filament undergoes such a well defined oscillation, the quantities  $\omega$ and $\delta$ are only fitted. 
Below, we give a more meaningful analytical interpretation. We first focus on $\omega$, as the $\delta$ is much smaller and therefore the frequency of oscillation only very weakly depends on $\delta$. To first order, the oscillation here should be only caused by gravity, as in our initial conditions the filament is at rest. In the following, we are going to calculate the force on a peak of the bend. We assume the cylinder to be a string, curved in the same way as in the simulations. For convenience, we shift to the cosine so that the peak is at the origin.
\begin{equation}
y = y_0 \thinspace cos(2\pi{\it W}\frac{x}{\it l}),
\end{equation}
where ${\it l}$ is the box length and ${\it W}$, the wavenumber, e.g. ${\it W}=3$ in our fiducial case. To calculate the gravitational force on a fluid parcel at the maximum at a current amplitude $y_t$ we determine the distance $r$ to each parcel along the string as well as the distance to each parcel in y-direction $\Delta y$
\begin{eqnarray}
\Delta y = y_t\left(1-cos(2\pi{\it W}\frac{x}{\it l})\right)\\
r = \sqrt{\Delta x^2 + \Delta y^2}\approx \Delta x = |x|.
\end{eqnarray}
Here, we approximated $r$ for small angles, i.e. $y_t<< {\it l}$. The gravitational acceleration in y-direction is then given by
\begin{equation}
a=\frac{-GM}{r^2}\frac{\Delta y}{r}=\int\frac{-G\left(\frac{M}{L}\right)_{\rm crit}f_{\rm cyl}\Delta y}{|x|^3}\thinspace dx,
\end{equation}
where we expressed the mass $M$ as integral over the line-mass. Substituting the variable ($u(x):=2\pi\frac{\it W}{\it l}x$) we arrive at
\begin{equation}
a = -G\left(\frac{M}{L}\right)_{\rm crit}f_{\rm cyl}\frac{4\pi^2{\it W}^2}{{\it l}^2}y_t\int\frac{1-cos(u)}{|u|^3}du.
\end{equation}
The integral diverges at 0, therefore we have to chose a different lower bound. A reasonable choice is the thickness, i.e. the radius R, of the filament, thereby assuming that within a filament the forces from above and below cancel\footnote{For the fiducial case the radius is close to $u(R)\approx\pi/2$ in terms of the cosine. This actually indicates that the downward pull is only exerted by material past the zero crossing in our model.}. Using this approximation we introduce for the fiducial case
\begin{equation}
I_{\rm R} = \int_{u(R)}^{\infty}\frac{1-cos(u)}{|u|^3}du\approx0.33\label{eq_integ}
\end{equation}
and using the symmetry arrive at the ODE
\begin{equation}
a = \ddot{y}(t) = I_{\rm R}\frac{-8\pi^2G\left(\frac{M}{L}\right)_{\rm crit}f_{\rm cyl}{\it W}^2}{{\it l}^2}y(t).
\end{equation}
This ODE has an analytic solution,
\begin{equation}
y(t) = y_0\thinspace {\rm cos}\left(\sqrt{I_{\rm R}\frac{8\pi^2G\left(\frac{M}{L}\right)_{\rm crit}f_{\rm cyl}{\it W}^2}{{\it l}^2}}t\right).\label{eq_yevol}
\end{equation}

\begin{figure}
\begin{center}
{\centering 
\includegraphics[width=8.5cm]{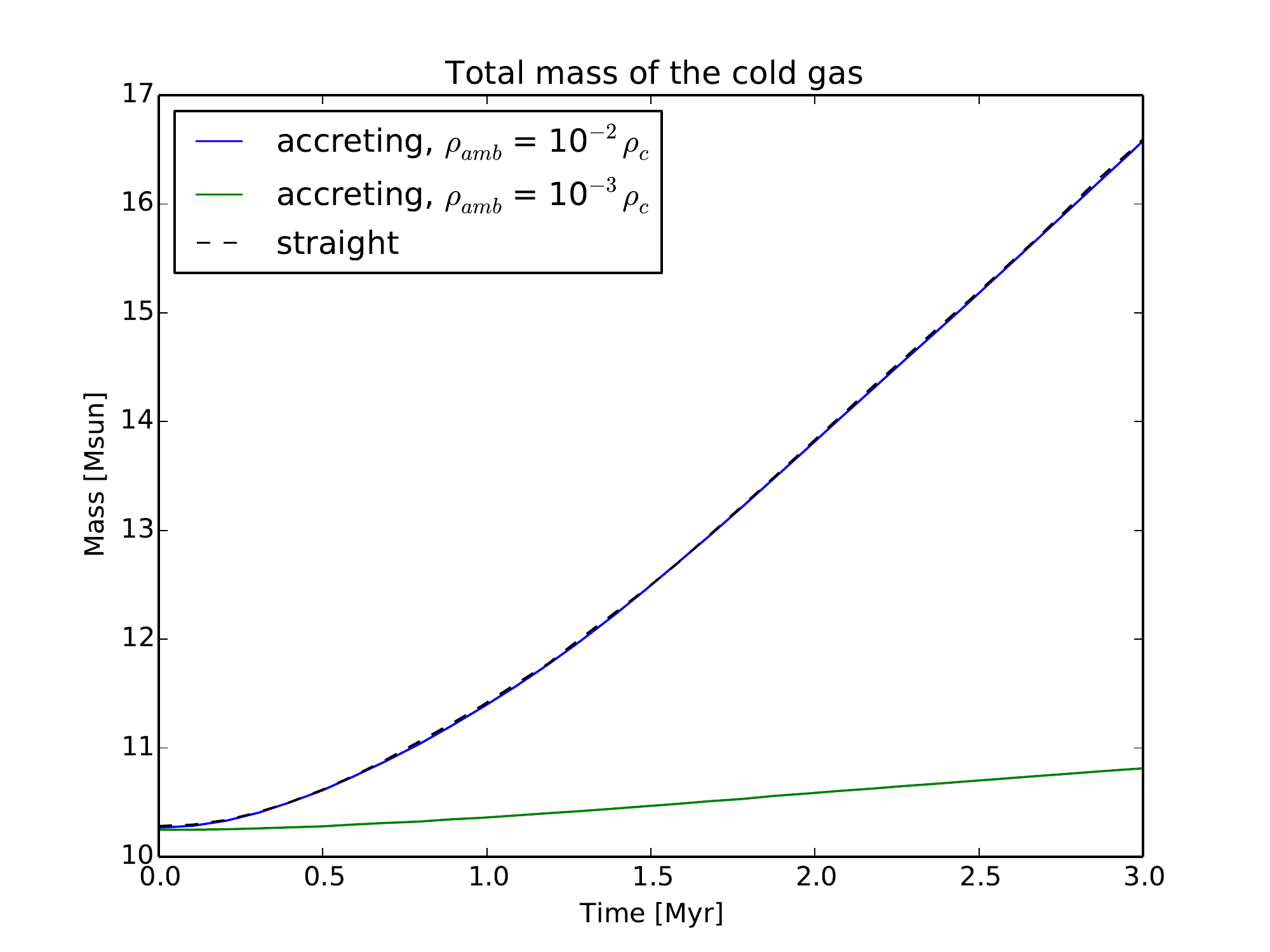}
}
\end{center}
\caption{Time evolution of the total mass of the cold gas in different cases with accretion. \label{fig:m_evol_accr}}
\end{figure}

\begin{figure}
\begin{center}
{\centering 
\includegraphics[width=8.5cm]{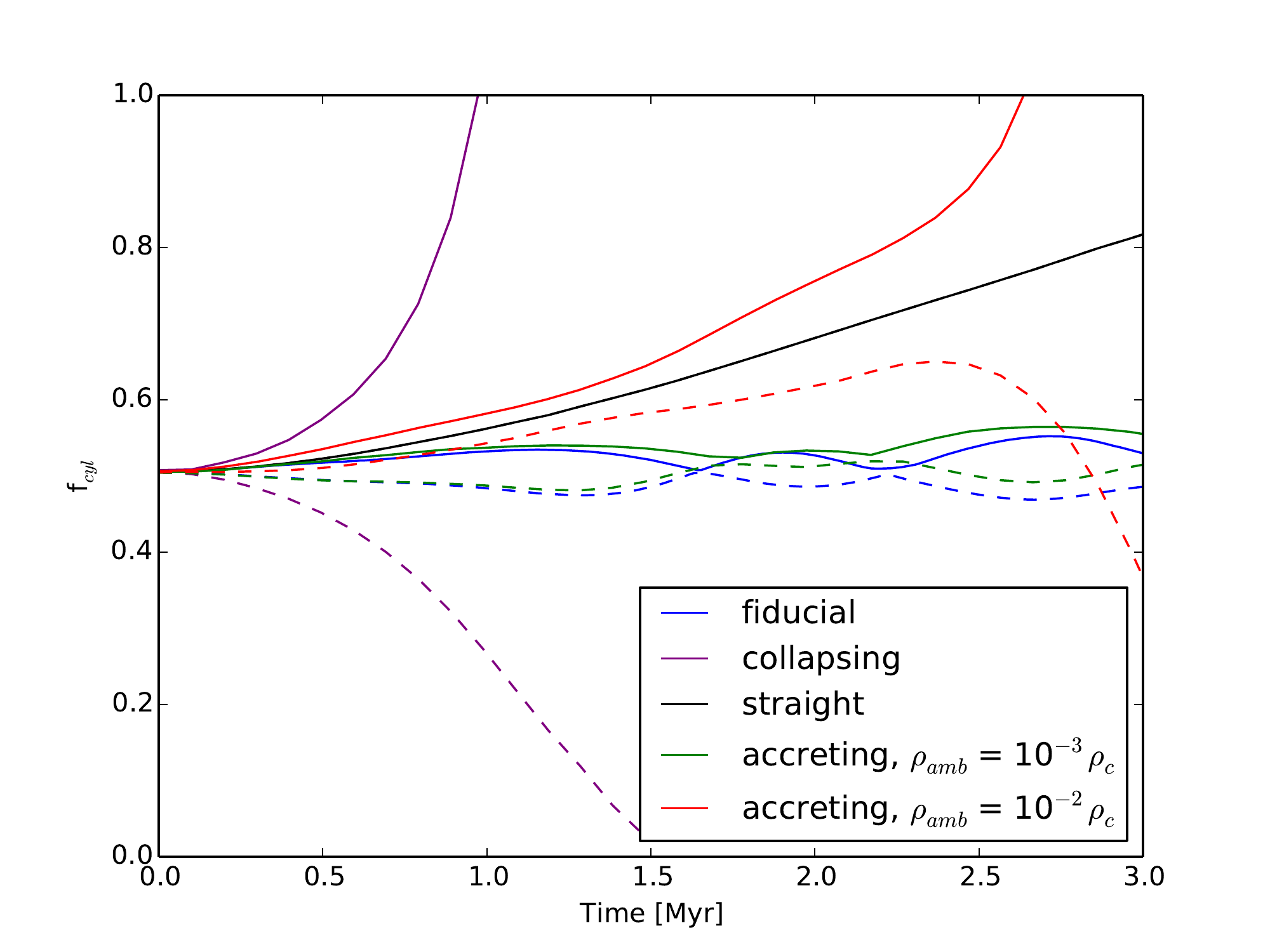}
}
\end{center}
\caption{Time evolution of the line mass. Solid line: maximum line mass, dashed lines: minimum line mass. \label{fig:f_evol}}
\end{figure}
\begin{figure}
\begin{center}
{\centering 
\includegraphics[width=8.5cm]{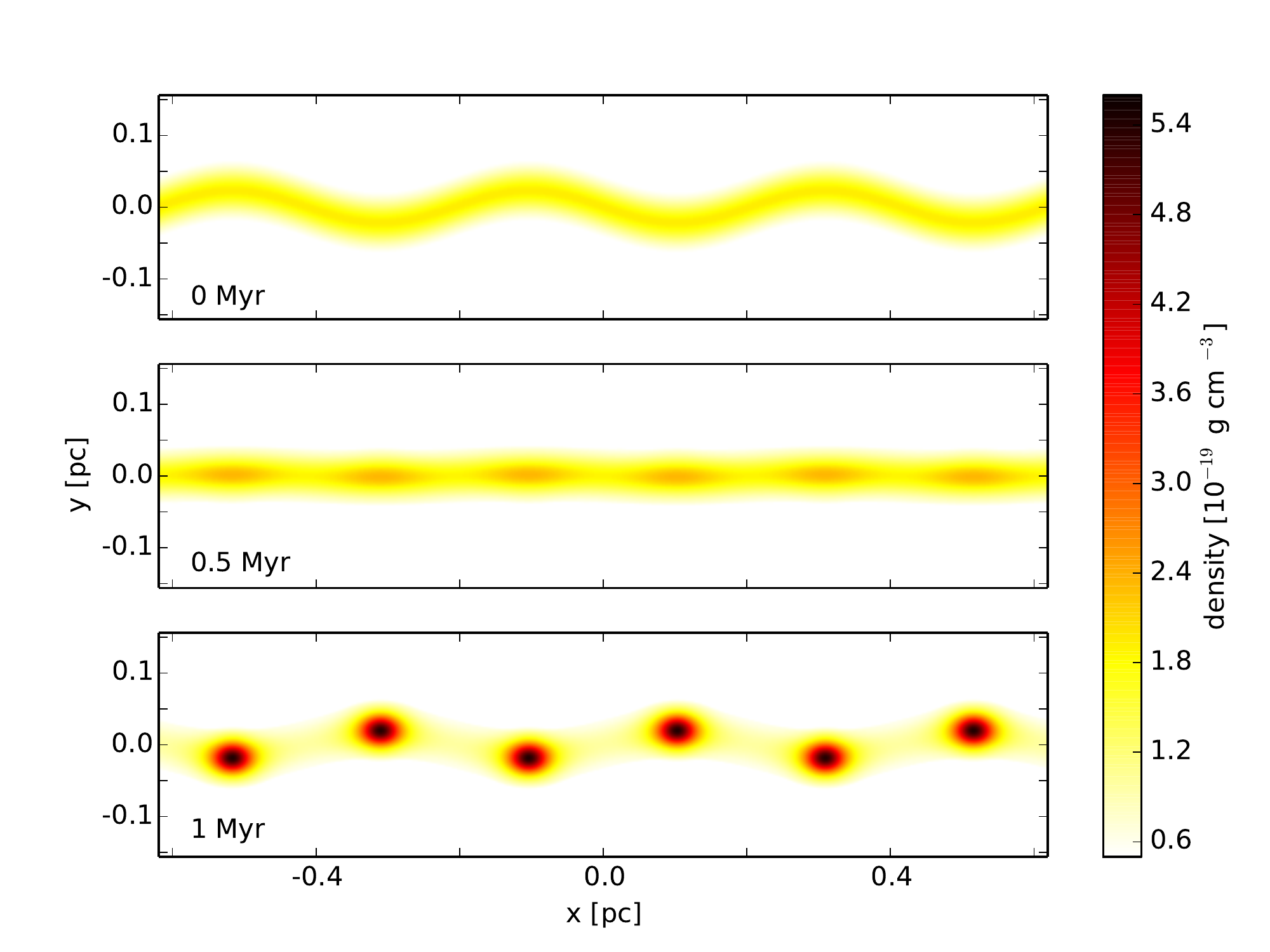}
}
\end{center}
\caption{Time evolution of the case with gravitational collapse. If the initial central density $n_c \ge 5\times 10^4\ndens$ the clumps forming in the simulations undergo gravitational collapse. After $1\Myr$ the simulation has turned into a string of oscillating cores/protostars. The spacing and number of the clumps is directly related to the initial perturbation. \label{fig:coll}}
\end{figure}
Therefore, the oscillation period is
\begin{equation}
T=\sqrt{\frac{{\it l}^2}{8\pi^2G\left(\frac{M}{L}\right)_{\rm crit}f_{\rm cyl}{\it W}^2 I_{\rm R}}}\label{eq_period}.
\end{equation}
We plot the solution as a green line in  Figure \ref{fig:sims_evol} (top panel). Despite a few approximations, it is a very good fit. 
Next, we will look into the damping. Presumably, the oscillation is damped by the inertia of the surrounding. In order to test this conjecture we reduce the density of the surrounding medium by a factor of 10 while at the same time increasing its temperature by a factor of 10 in order to guarantee pressure equilibrium. The result is shown in Fig \ref{fig:sims_evol} (bottom panel). The blue line is our fiducial simulation, the green line is the new test. When the ambient density is reduced, the damping vanishes and our analytical solution is in better agreement with the numerical result. From this, we conclude that the damping is caused by momentum transfer of the oscillating filament to the surrounding. This is remarkable, as it means that an oscillation is damped more strongly in a denser environment, independent of the thermal pressure in the surrounding. As a consequence, this indicates that filaments in dense regions should be more straight. In environments with $n\approx 100\ndens$ like in our fiducial run the filaments should be less bent than filaments surrounded by more disperse gas. 

As an additional test, we performed a simulation with the fiducial density contrast but increased the amplitude $y_0$ to $4\%$. The rescaled result is shown as black dashed line in  Figure \ref{fig:sims_evol} (bottom panel). The frequency is unaffected, as expected - $\omega$ is independent of $y_0$ - but the damping is twice as strong, as the path is twice as long.
\subsection{Accretion}
Another aspect is the potential ongoing accretion onto a filament, which is sometimes observed \citep[e.g.][]{Palmeirim:2013}. In the simulations above, no accretion takes place as the ambient gas is not allowed to cool. In order to allow for very simplified cooling we set the ambient medium to the temperature of the cold gas as soon as it reaches the border density of the filament. The resulting accretion is of course strongly dependent on the density of the ambient medium. In our fiducial case, the accretion increases the line mass up to the critical value within $3\Myr$. As can been seen in Fig. \ref{fig:m_evol_accr}, the accretion is not affected by the geometry of the filament as a straight filament accretes on the same time-scale. If the ambient medium is an order of magnitude less dense, the accretion is so weak the the fiducial case is essentially recovered. The oscillation frequency is only weakly affected. In the case with significant accretion the filament oscillates slightly faster, as it is more massive. However, the fragmentation is affected, as the critical value for $f_{\rm cyl}$ is reached earlier (see below, \S \ref{fragmentation}).
\subsection{Further Parameter Study}
\label{furtherparam}
We compare the above estimates with additional simulations.  Figure \ref{fig:sims_ps} (top panel) shows simulations with three periods of the sine and a varying mass-to-line ratio, parametrized by $f_{\rm cyl}$. 
We over-plot the solutions according to Eq. \ref{eq_yevol} with dashed lines. In general, the dependence follows $\sqrt{f_{\rm cyl}}$, as expected from the derivation above. The cases for $f_{\rm cyl}=0.95$ and $f_{\rm cyl}=0.5$ are in complete agreement. 
The case of $f_{\rm cyl}=0.25$ contains a different initial density, this is indicated by the asterisk in the legend. To understand this it is important to keep in mind that we kept the central density, $\rho_c$, fixed, the different values of $f_{\rm cyl}$ are realized by cutting off the profile at different positions. Therefore, a filament with a higher $f_{\rm cyl}$ will simply be wider. The filament with $f_{\rm cyl} = 0.25$ is the thinest in our study and very similar to a constant density, as only the central part of the profile is cut out. It is only half as wide as the other cases, therefore the chosen value of the lower bound of the integration in $I_{\rm R}$ (Eq. \ref{eq_integ}) is too big. As a result, the filament would oscillate faster than predicted. To make the situation more comparable, we halved the central density to $n_c = 5\times 10^3\ndens$. As a result, the width is now comparable to the other cases and the filament oscillates as predicted (Figure \ref{fig:sims_ps}, top panel, black lines). Another option would have been to re-evaluate $I_{\rm R}$, but as this test shows, this is unnecessary, as long as $R\gtrsim0.15\pc$. Altogether, the equations derived above describe the evolution of the filament very well in all three cases.

In  Figure \ref{fig:sims_ps} (bottom panel) we test different wave-numbers, i.e a different wave-length, while keeping $f_{\rm cyl}=0.5$. These simulations are performed with a box length of $1\pc$. Again, we over-plot the solutions according to Eq. \ref{eq_yevol} with dashed lines. As expected, the oscillation is faster for simulations with shorter wave-length. The very good agreement for ${\it W}=3$ for a different box length than before shows the robustness of our estimates. The case for ${\it W}=2$ is close to the predictions as well. However, the high wave number case ${\it W}=5$ oscillates much slower than expected. This is due to the amplitude getting comparable to the wave-length. In this case, the analytical small angle approximation becomes worse. It is only a small discrepancy in the frequency but over the course of $2.5\Myr$ it adds up. 
\begin{figure}
\begin{center}
{\centering 
\includegraphics[width=4.2cm]{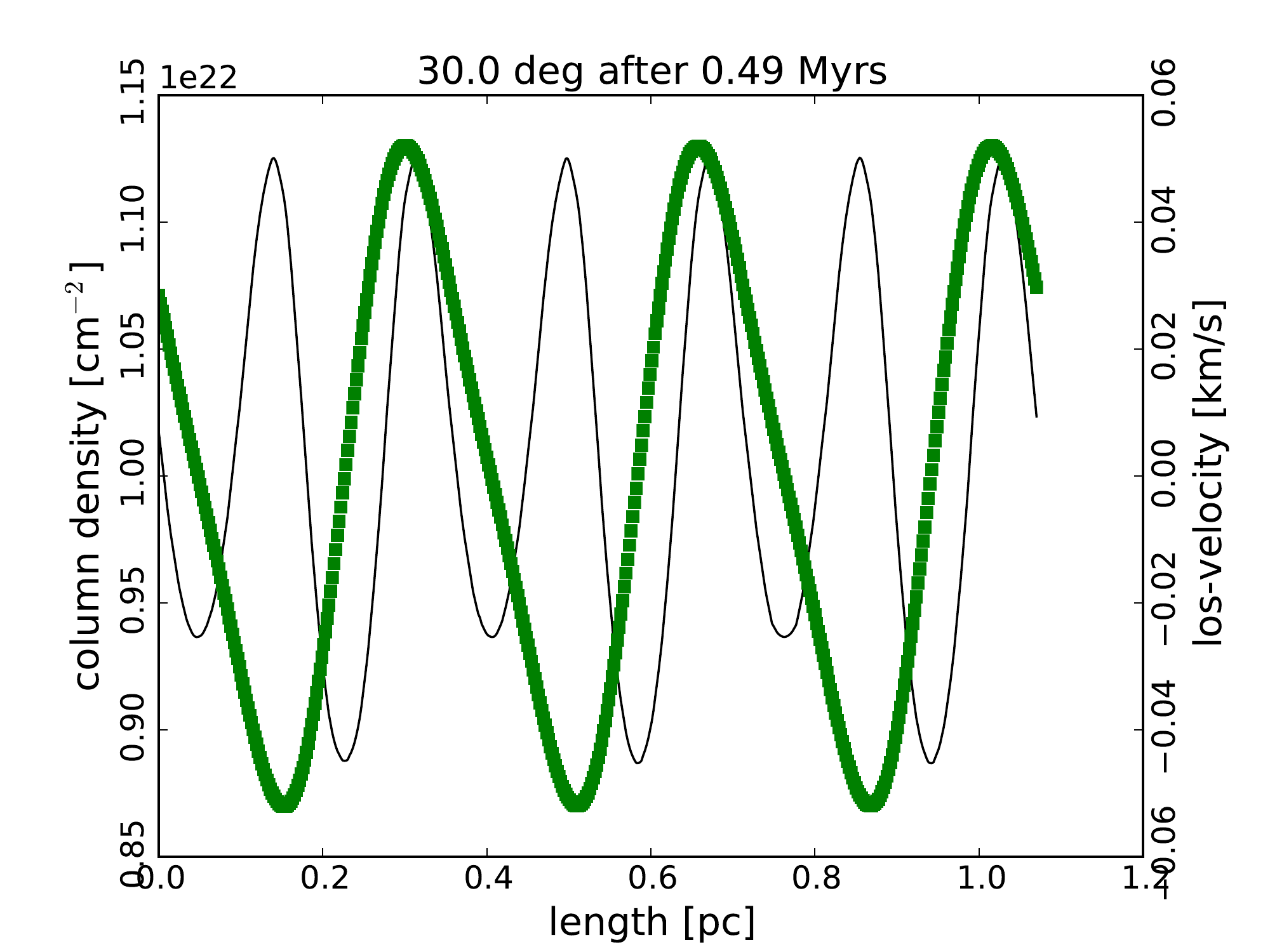}
\includegraphics[width=4.2cm]{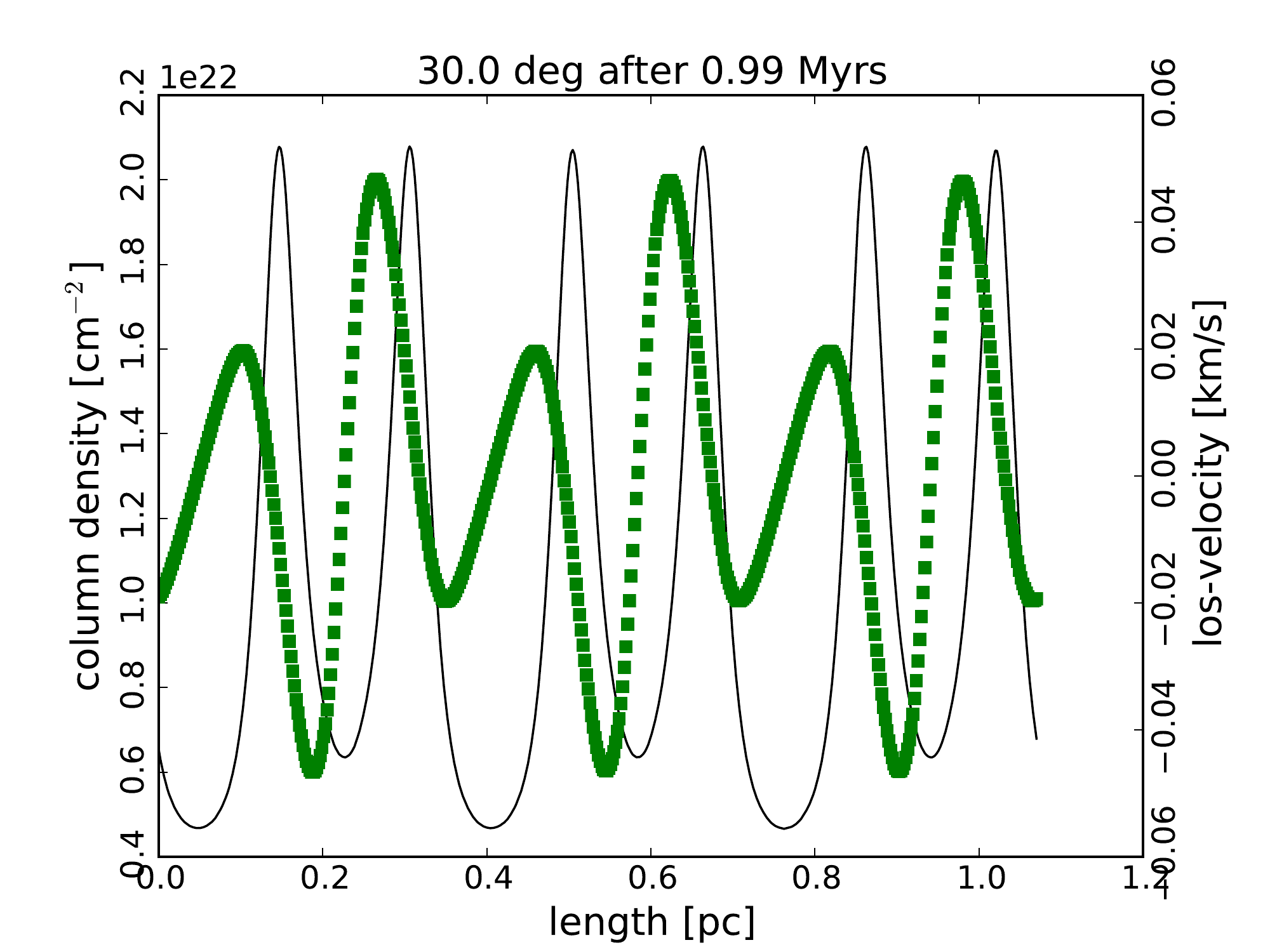}
\includegraphics[width=4.2cm]{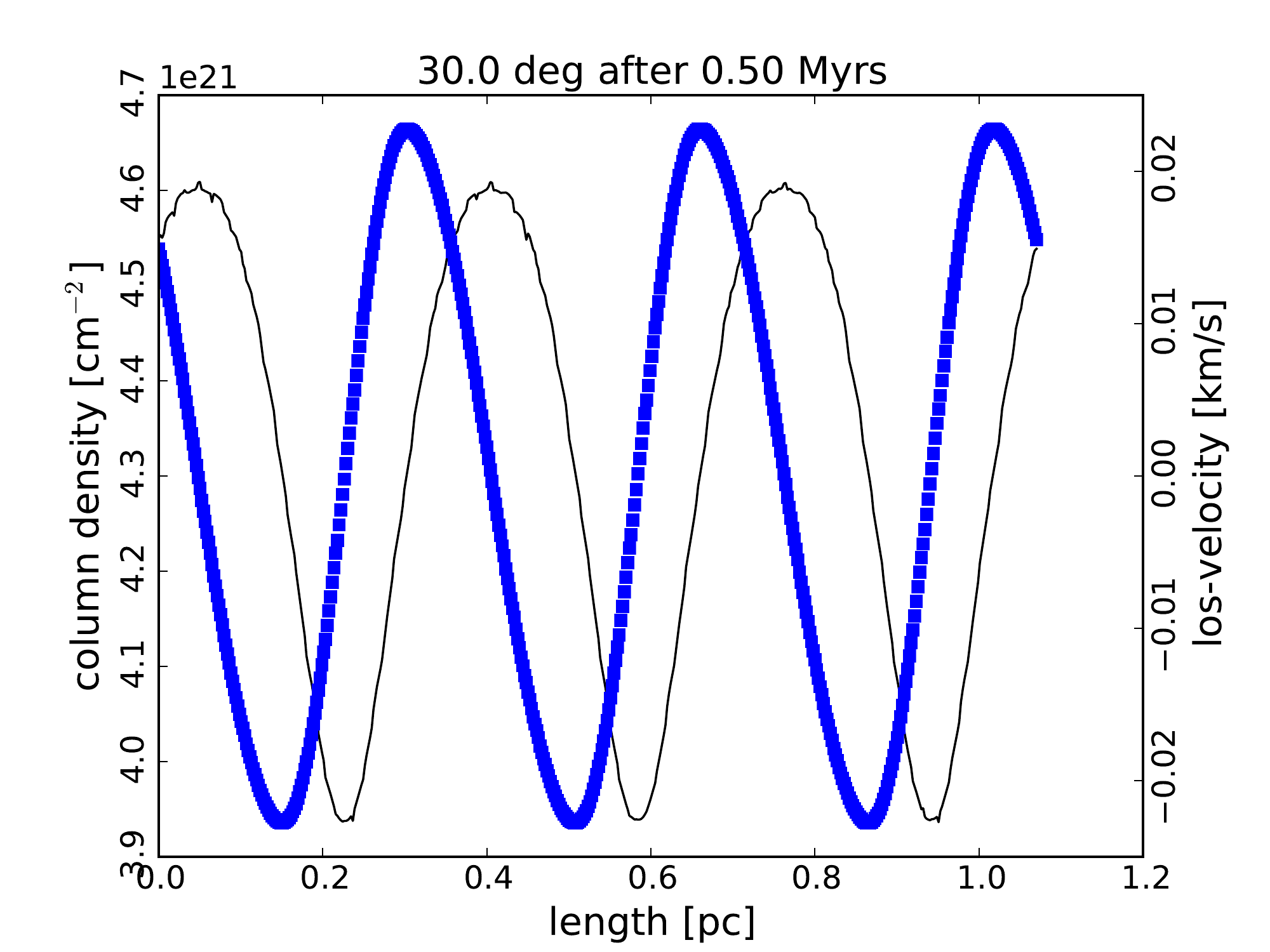}
\includegraphics[width=4.2cm]{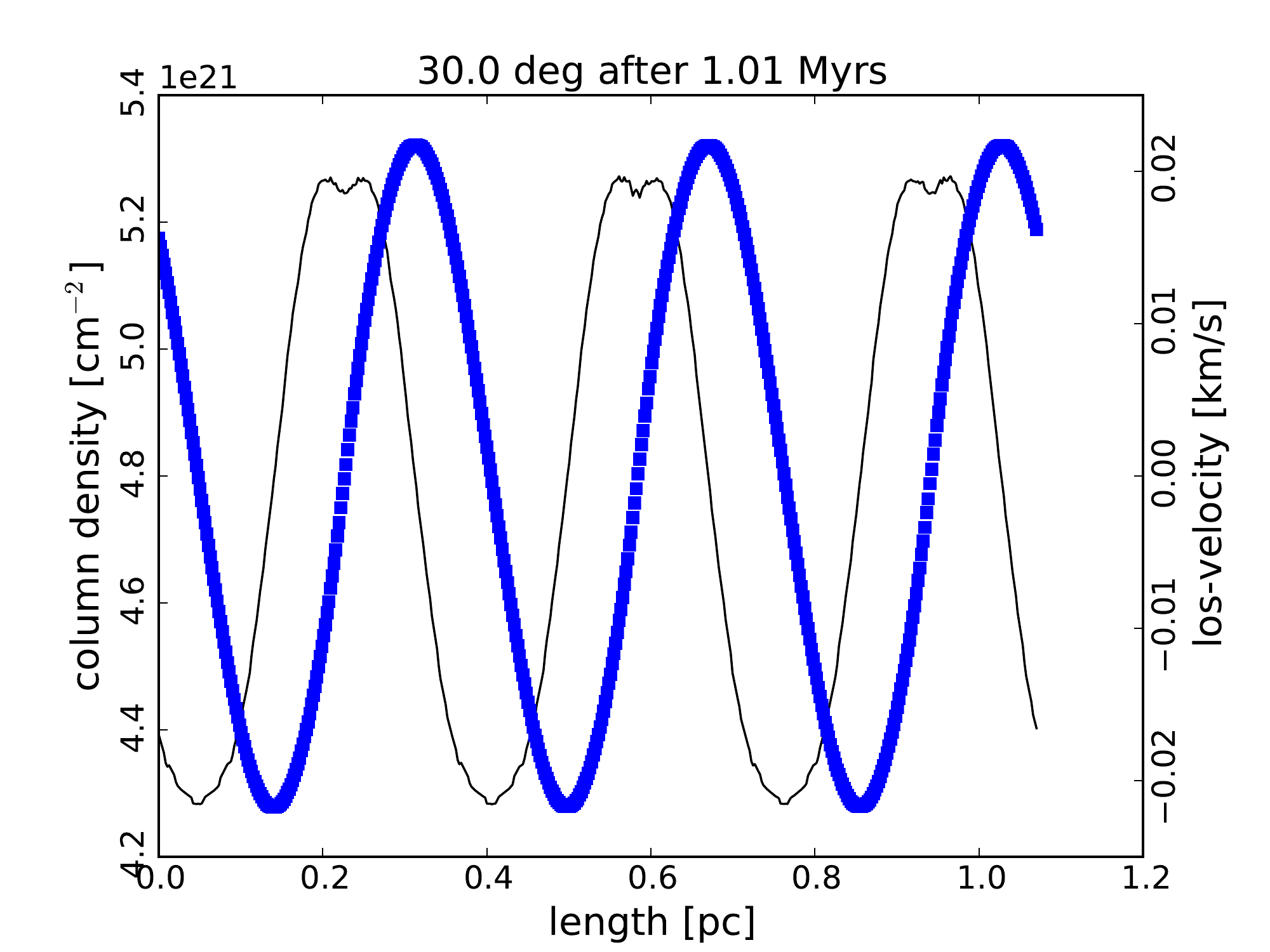}
\includegraphics[width=4.2cm]{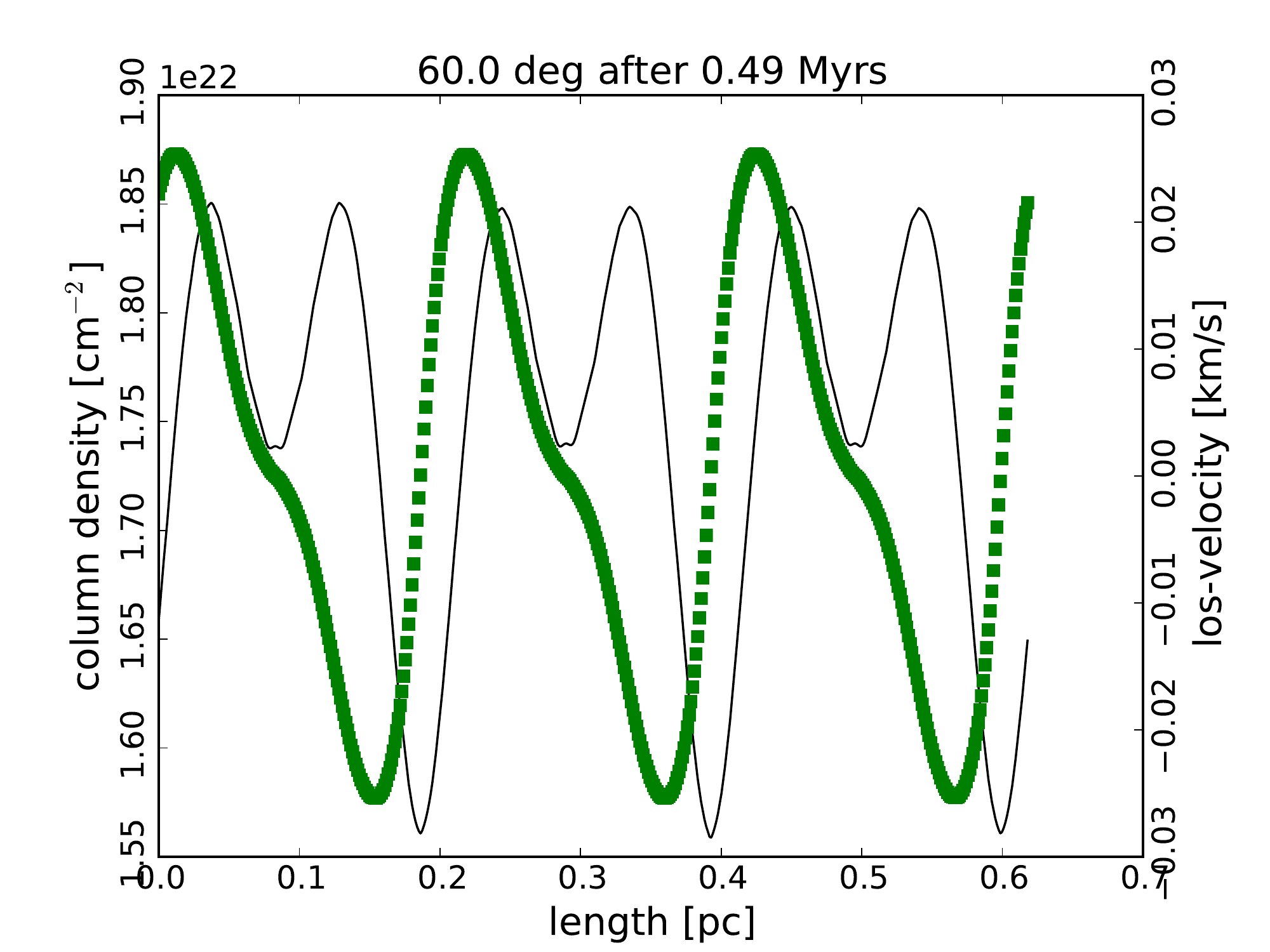}
\includegraphics[width=4.2cm]{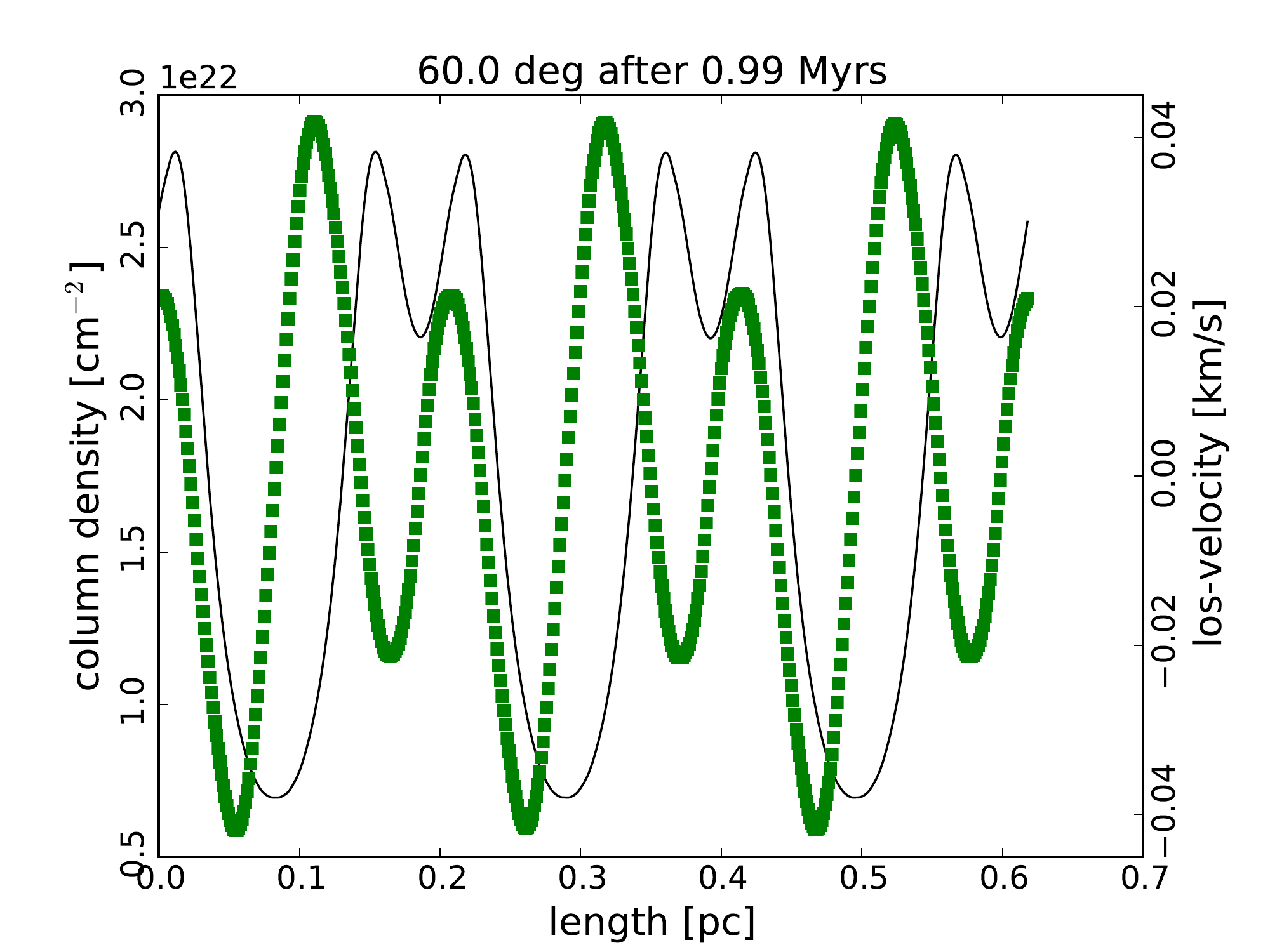}
\includegraphics[width=4.2cm]{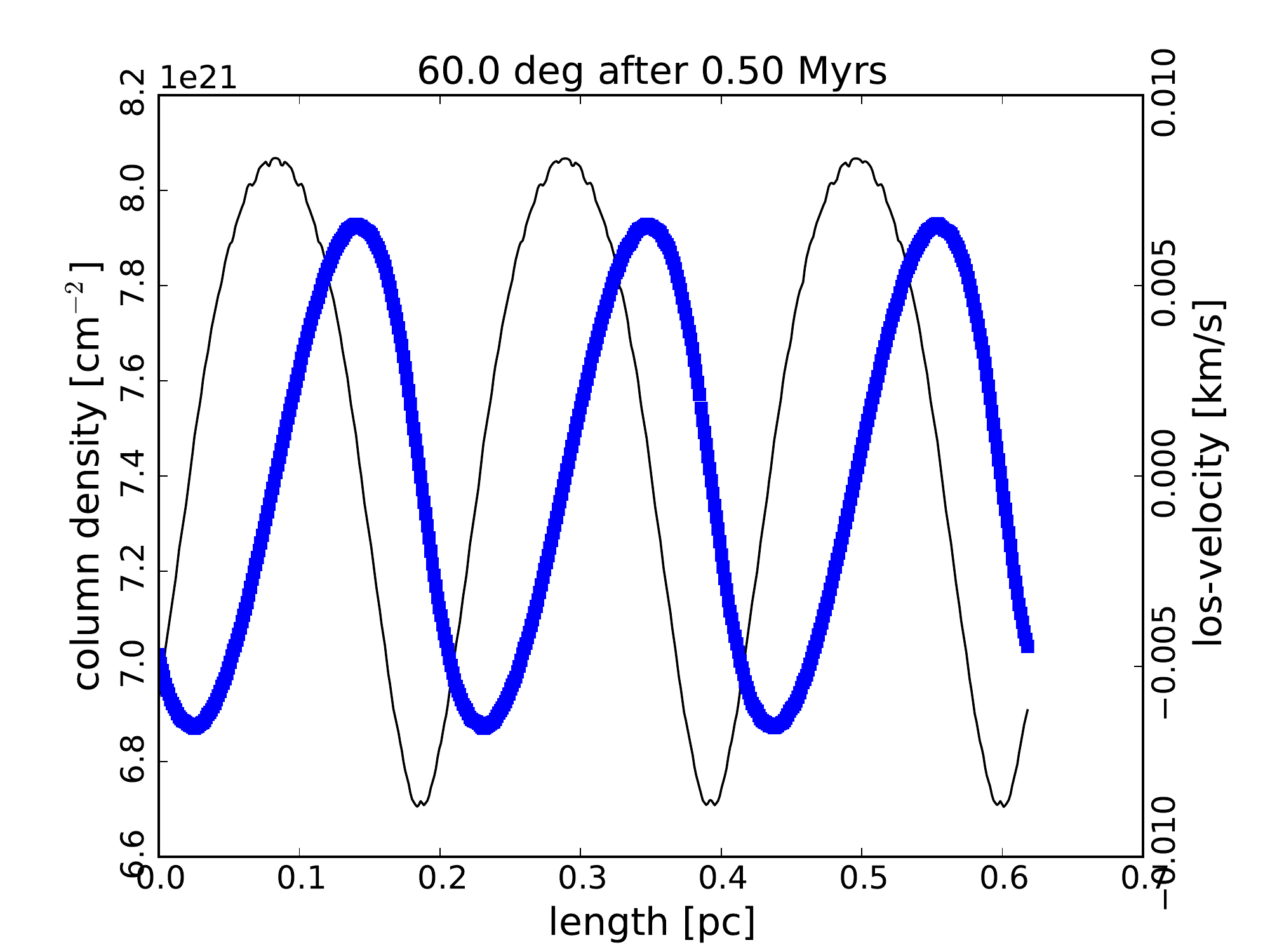}
\includegraphics[width=4.2cm]{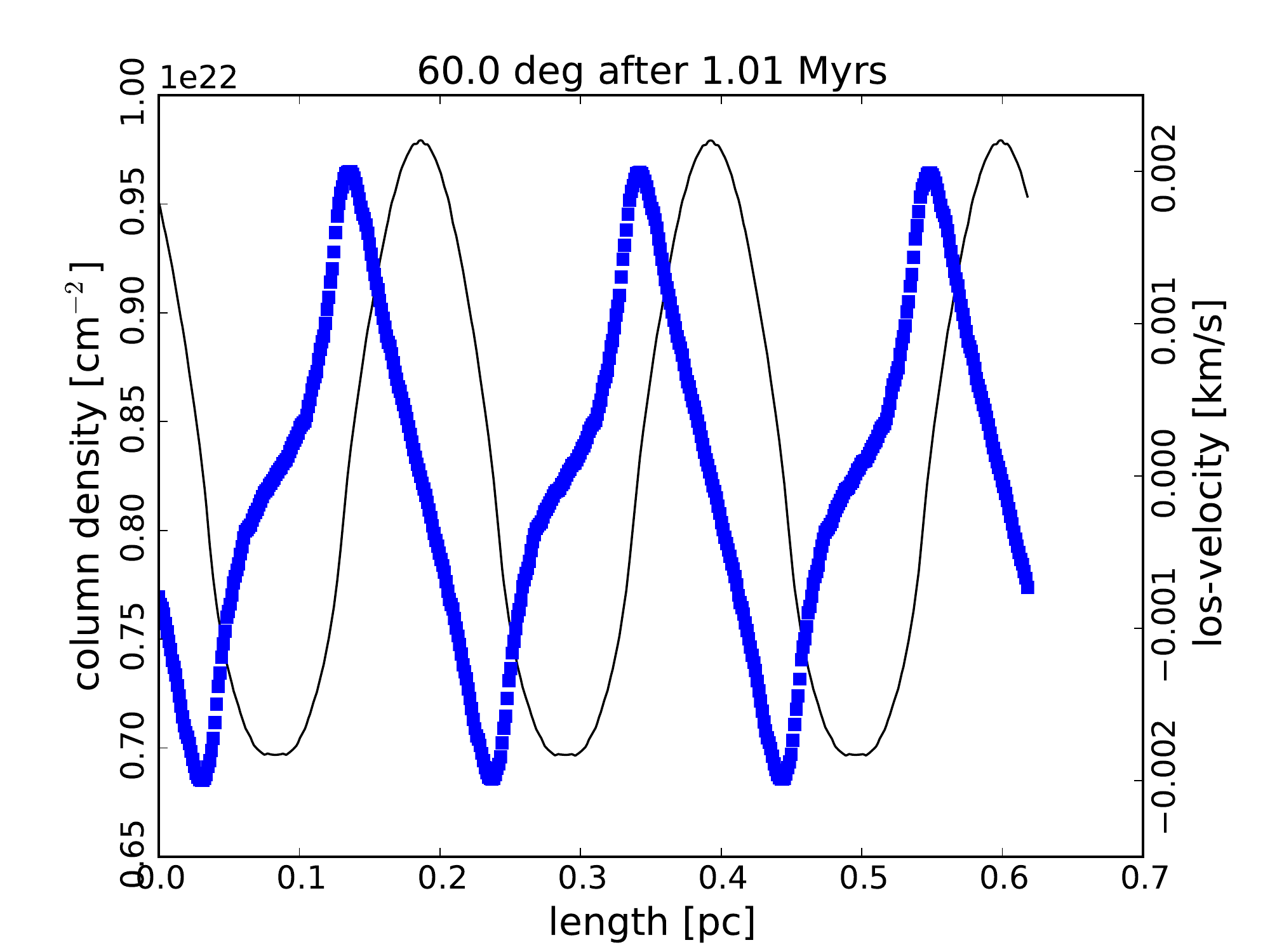}
}
\end{center}
\caption{Time evolution for mock observations. We calculate the line profiles for the velocities as they would be observed using $C^{18}O$ as a tracer. Left columns is $0.5\Myr$, right is $1\Myr$. The velocity centroids are shown as blue boxes for the fiducial simulations from Figure \ref{fig:sims_osc} and as green boxes for the collapsing case from Figure \ref{fig:coll}. The underlying column density is over-plotted as black lines. The top four panels are at an inclination of $30^\circ$, the bottom four panels at $60^\circ$.  \label{fig:coll_vel}}
\end{figure}
\subsection{Fragmentation}
\label{fragmentation}
The last aspect we want to focus on is the fragmentation of bending filaments. As seen in  Figure \ref{fig:sims_osc}, denser regions are formed at the positions of the initial maxima and minima of the geometry perturbation. These regions form at the turning point of the oscillation, when the velocities are lowest and the material gets piled up. Their density is enhanced by a factor of 2-3. However, they are transient and disappear only to re-appear at the next oscillation maximum. Therefore, our fiducial filament does not fragment within the time simulated here ($\approx 5\Myr$). The line mass fluctuates a bit but stays in the stable regime $f_{\rm cyl} \approx 0.45-0.55$ (see Fig. \ref{fig:f_evol}, blue lines). 

To investigate potential fragmentation, we performed another set of simulations with increased initial central density $\rho_{\rm c}$. We find that for $n_c=5\times10^4\ndens$ and above the enhanced regions collapse gravitationally. This can be seen in  Figure \ref{fig:coll}, where we show the time evolution for $f_{\rm cyl} = 0.5$ and $n_c=5\times10^4\ndens$. The initial subcritical and stable configuration starts to oscillate and after $\approx 1\Myr$ the pile-up of material leads to gravitational run-away collapse and a line-mass $f_{\rm cyl} > 1$ (see Fig. \ref{fig:f_evol}, purple lines). The locations of the cores are exactly aligned with the initial perturbation and the filament transforms into a string of oscillating cores and later most likely stars. We call this process `geometrical fragmentation', because the locations of the collapse are given by the geometry of the filament or perturbations. Still, gravity is of course the driving force behind the oscillation as well as the fragmentation.

In the case with accretion, we have to distinguish between different densities of the ambient medium. For the low density ambient medium ($\rho_{\rm amb} = 10^{-3}\rho_{\rm c}$) the accretion is not very effective and the scenario is the same as in the fiducial case. The line mass fluctuates, but stays in the stable regime (Fig. \ref{fig:f_evol}, green lines). In the case with a denser ambient medium ($\rho_{\rm amb} = 10^{-2}\rho_{\rm c}$) the situation changes. The oscillations are still visible but on top of the general increase in the line mass due to the accretion (Fig. \ref{fig:f_evol}, red lines). After $~2.25\Myr$  the critical value is reached and the filament collapses. Still, it fragments into six fragments, at exactly the postions of the maxima and minima of the geometric perturbation. Therefore, we consider this mode of collapse to be geometrical fragmentation as well. 

It is important to keep in mind that these simulations are performed without any initial density perturbations. With density perturbations, a competition between the `geometrical fragmentation' discussed here and gravitational fragmentation is to be expected. It will depend on the relative time scales and length scales, which one prevails. As mentioned before, our box length is set to two times the length of the fastest growing mode for a density perturbation. Once we introduce this perturbation there will be two gravitational and six geometrical density enhancements competing for the material in the filament. We will discuss this complex situation in a future publication in this series.
\subsection{Observed velocities}
To facilitate an easier comparison with observations and to distinguish between collapse and oscillations we next analyzed the observable line-of-sight (LOS) velocities.
The velocities are calculated as in \citet{Heigl:2016}. Each volume element of the simulation is assumed to emit a discrete LOS value. This value is then converted into a Gaussian line profile with $\sigma=0.0526\kms$, corresponding to the thermal line-width of ${\rm C^{18}O}$, and density weighted. The profiles obtained in this way are then sorted into bins of $0.05\kms$. The velocity centroid is measured by fitting a Gaussian to the line, as it is done for observations.
The filaments are observed under two inclination angles, $30^\circ$ and $60^\circ$. We show a few examples in Figure \ref{fig:coll_vel}. The left column is at $0.5\Myr$, the right column at $1.0\Myr$. We show the velocities for the collapsing case from Figure \ref{fig:coll} in green and for the fiducial case from Figure \ref{fig:sims_osc} in blue. The black lines show the column density distribution. The top four panels are at an inclination of $30^\circ$, the bottom four panels at $60^\circ$.

The first striking result is that already our values for oscillating filaments without collapse (blue) are comparable to e.g. the $\delta v_{\rm lsr}\approx\pm0.05\kms$ observed for a fragmenting filament in L1517 by \citep{Hacar:2011}. However, the profiles for the fiducial case are not indicative of collapse.
The collapsing case (green lines) shows a wing shaped profile at $1\Myr$, the classic sign of infall (second column, rows one and three, respectively), whereas the fiducial case is still stable. At the earlier stages there are no obvious signs of infall. Still, there is a remarkable feature: the six peaks in density correspond to only tree peaks in velocity - the other peaks are hidden by the oscillation velocities. This allows to explain peculiar observations like the ones in the in the Taurus B213 region in the L1495/B213 filament \citep{Tafalla:2015aa}, where the velocities do not seem to coincide with the surface density. Our suggestion is that they are inclined bent filaments in the process of fragmentation.\vspace{1cm}

\section{Summary \& Conclusions}
\label{summary}
We treat a filament as a cylinder of infinite length in hydrostatic equilibrium. We find that a slight initial bend of a filament greatly influences its evolution. The gravitational pull of the different regions onto each other leads to a long lived, stable oscillation. Deriving analytical estimates and comparing them with a parameter study we are able to show that the oscillation period mainly depends on the line-mass and the initial wavenumber or wavelength of the perturbation. At the turning time of the oscillation, the gas gets piled up at the position of the maxima and minima and leads to density enhancements. In our fiducial case, these enhancements are transient. However,  
 in the denser cases ($n_c \ge 5\times 10^4\ndens$) the enhancements become gravitational unstable and their line-mass exceeds the critical value. Another process that can lead to fragmentation is accretion onto the filament. When the ambient gas is allowed to cool as soon as it reaches the density of the filament, the accretion of this cool gas leads to an increase in the line-mass. After $2.25\Myr$ the critical value is exceeded and fragmentation occurs. Again, the filament fragments on the length scale of the intial geometrical perturbation.
Either effect allows to fragment a filament on any given length-scale. To be able to compare this new evolutionary state of filaments to observations we investigate the LOS velocities. The values agree with observations. Under certain angles, the LOS velocities show signs of infall onto the correct number of cores. However, under higher inclination angles (e.g. $60^\circ$)  half of the cores are hidden in the velocity information due to the overlay of the velocities of the oscillation with the velocities of the fragmentation.  The cores can only be disentangled in the column density or intensity - if the resolution is high enough.

Altogether, the geometrical shape of a filament greatly influences clump and therefore core formation as well as the observables.
We call the newly found process described above `geometrical fragmentation'.

\section{Acknowledgements}
MG acknowledges funding by the DFG priority program 1573 `Physics of the Interstellar Medium'.

\bibliographystyle{apj}

\begin{thebibliography}{29}
\expandafter\ifx\csname natexlab\endcsname\relax\def\natexlab#1{#1}\fi

\bibitem[{{Andr{\'e}} {et~al.}(2010){Andr{\'e}}, {Men'shchikov}, {Bontemps},
  {K{\"o}nyves}, {Motte}, {Schneider}, {Didelon}, {Minier}, {Saraceno},
  {Ward-Thompson}, {di Francesco}, {White}, {Molinari}, {Testi}, {Abergel},
  {Griffin}, {Henning}, {Royer}, {Mer{\'{\i}}n}, {Vavrek}, {Attard},
  {Arzoumanian}, {Wilson}, {Ade}, {Aussel}, {Baluteau}, {Benedettini},
  {Bernard}, {Blommaert}, {Cambr{\'e}sy}, {Cox}, {di Giorgio}, {Hargrave},
  {Hennemann}, {Huang}, {Kirk}, {Krause}, {Launhardt}, {Leeks}, {Le Pennec},
  {Li}, {Martin}, {Maury}, {Olofsson}, {Omont}, {Peretto}, {Pezzuto}, {Prusti},
  {Roussel}, {Russeil}, {Sauvage}, {Sibthorpe}, {Sicilia-Aguilar}, {Spinoglio},
  {Waelkens}, {Woodcraft}, \& {Zavagno}}]{Andre:2010}
{Andr{\'e}}, P. {et~al.} 2010, \aap, 518, L102

\bibitem[{{Arzoumanian} {et~al.}(2011){Arzoumanian}, {Andr{\'e}}, {Didelon},
  {K{\"o}nyves}, {Schneider}, {Men'shchikov}, {Sousbie}, {Zavagno}, {Bontemps},
  {di Francesco}, {Griffin}, {Hennemann}, {Hill}, {Kirk}, {Martin}, {Minier},
  {Molinari}, {Motte}, {Peretto}, {Pezzuto}, {Spinoglio}, {Ward-Thompson},
  {White}, \& {Wilson}}]{Arzoumanian:2011}
{Arzoumanian}, D. {et~al.} 2011, \aap, 529, L6

\bibitem[{{Burkert} \& {Hartmann}(2004)}]{Burkert:2004}
{Burkert}, A., \& {Hartmann}, L. 2004, \apj, 616, 288

\bibitem[{{Clarke} \& {Whitworth}(2015)}]{Clarke:2015}
{Clarke}, S.~D., \& {Whitworth}, A.~P. 2015, \mnras, 449, 1819

\bibitem[{{Federrath}(2015)}]{Federrath:2015aa}
{Federrath}, C. 2015, ArXiv e-prints

\bibitem[{{Field}(1965)}]{Field:1965}
{Field}, G.~B. 1965, \apj, 142, 531

\bibitem[{{Fischera} \& {Martin}(2012)}]{Fischera:2012aa}
{Fischera}, J., \& {Martin}, P.~G. 2012, \aap, 542, A77

\bibitem[{{Hacar} \& {Tafalla}(2011)}]{Hacar:2011}
{Hacar}, A., \& {Tafalla}, M. 2011, \aap, 533, A34

\bibitem[{{Hacar} {et~al.}(2013){Hacar}, {Tafalla}, {Kauffmann}, \&
  {Kov{\'a}cs}}]{Hacar:2013}
{Hacar}, A., {Tafalla}, M., {Kauffmann}, J., \& {Kov{\'a}cs}, A. 2013, \aap,
  554, A55

\bibitem[{{Hartmann}(2002)}]{Hartmann:2002}
{Hartmann}, L. 2002, \apj, 578, 914

\bibitem[{{Heigl} {et~al.}(2016){Heigl}, {Burkert}, \& {Hacar}}]{Heigl:2016}
{Heigl}, S., {Burkert}, A., \& {Hacar}, A. 2016, ArXiv e-prints

\bibitem[{{Heitsch}(2013{\natexlab{a}})}]{Heitsch:2013a}
{Heitsch}, F. 2013{\natexlab{a}}, \apj, 769, 115

\bibitem[{{Heitsch}(2013{\natexlab{b}})}]{Heitsch:2013b}
---. 2013{\natexlab{b}}, \apj, 776, 62

\bibitem[{{Heitsch} {et~al.}(2006){Heitsch}, {Slyz}, {Devriendt}, {Hartmann},
  \& {Burkert}}]{Heitsch:2006}
{Heitsch}, F., {Slyz}, A.~D., {Devriendt}, J.~E.~G., {Hartmann}, L.~W., \&
  {Burkert}, A. 2006, \apj, 648, 1052

\bibitem[{{Keto} \& {Burkert}(2014)}]{Keto:2014}
{Keto}, E., \& {Burkert}, A. 2014, \mnras, 441, 1468

\bibitem[{{Lada} {et~al.}(2008){Lada}, {Muench}, {Rathborne}, {Alves}, \&
  {Lombardi}}]{Lada:2008}
{Lada}, C.~J., {Muench}, A.~A., {Rathborne}, J., {Alves}, J.~F., \& {Lombardi},
  M. 2008, \apj, 672, 410

\bibitem[{{Moeckel} \& {Burkert}(2015)}]{Moeckel:2015aa}
{Moeckel}, N., \& {Burkert}, A. 2015, \apj, 807, 67

\bibitem[{{Molinari} {et~al.}(2010){Molinari}, {Swinyard}, {Bally}, {Barlow},
  {Bernard}, {Martin}, {Moore}, {Noriega-Crespo}, {Plume}, {Testi}, {Zavagno},
  {Abergel}, {Ali}, {Anderson}, {Andr{\'e}}, {Baluteau}, {Battersby},
  {Beltr{\'a}n}, {Benedettini}, {Billot}, {Blommaert}, {Bontemps}, {Boulanger},
  {Brand}, {Brunt}, {Burton}, {Calzoletti}, {Carey}, {Caselli}, {Cesaroni},
  {Cernicharo}, {Chakrabarti}, {Chrysostomou}, {Cohen}, {Compiegne}, {de
  Bernardis}, {de Gasperis}, {di Giorgio}, {Elia}, {Faustini}, {Flagey},
  {Fukui}, {Fuller}, {Ganga}, {Garcia-Lario}, {Glenn}, {Goldsmith}, {Griffin},
  {Hoare}, {Huang}, {Ikhenaode}, {Joblin}, {Joncas}, {Juvela}, {Kirk},
  {Lagache}, {Li}, {Lim}, {Lord}, {Marengo}, {Marshall}, {Masi}, {Massi},
  {Matsuura}, {Minier}, {Miville-Desch{\^e}nes}, {Montier}, {Morgan}, {Motte},
  {Mottram}, {M{\"u}ller}, {Natoli}, {Neves}, {Olmi}, {Paladini}, {Paradis},
  {Parsons}, {Peretto}, {Pestalozzi}, {Pezzuto}, {Piacentini}, {Piazzo},
  {Polychroni}, {Pomar{\`e}s}, {Popescu}, {Reach}, {Ristorcelli}, {Robitaille},
  {Robitaille}, {Rod{\'o}n}, {Roy}, {Royer}, {Russeil}, {Saraceno}, {Sauvage},
  {Schilke}, {Schisano}, {Schneider}, {Schuller}, {Schulz}, {Sibthorpe},
  {Smith}, {Smith}, {Spinoglio}, {Stamatellos}, {Strafella}, {Stringfellow},
  {Sturm}, {Taylor}, {Thompson}, {Traficante}, {Tuffs}, {Umana}, {Valenziano},
  {Vavrek}, {Veneziani}, {Viti}, {Waelkens}, {Ward-Thompson}, {White},
  {Wilcock}, {Wyrowski}, {Yorke}, \& {Zhang}}]{Molinari:2010}
{Molinari}, S. {et~al.} 2010, \aap, 518, L100

\bibitem[{{Ostriker} {et~al.}(2010){Ostriker}, {McKee}, \&
  {Leroy}}]{Ostriker:2010}
{Ostriker}, E.~C., {McKee}, C.~F., \& {Leroy}, A.~K. 2010, \apj, 721, 975

\bibitem[{{Ostriker}(1964)}]{Ostriker:1964aa}
{Ostriker}, J. 1964, \apj, 140, 1529

\bibitem[{{Palmeirim} {et~al.}(2013){Palmeirim}, {Andr{\'e}}, {Kirk},
  {Ward-Thompson}, {Arzoumanian}, {K{\"o}nyves}, {Didelon}, {Schneider},
  {Benedettini}, {Bontemps}, {Di Francesco}, {Elia}, {Griffin}, {Hennemann},
  {Hill}, {Martin}, {Men'shchikov}, {Molinari}, {Motte}, {Nguyen Luong},
  {Nutter}, {Peretto}, {Pezzuto}, {Roy}, {Rygl}, {Spinoglio}, \&
  {White}}]{Palmeirim:2013}
{Palmeirim}, P. {et~al.} 2013, \aap, 550, A38

\bibitem[{{Pon} {et~al.}(2012){Pon}, {Toal{\'a}}, {Johnstone},
  {V{\'a}zquez-Semadeni}, {Heitsch}, \& {G{\'o}mez}}]{Pon:2012}
{Pon}, A., {Toal{\'a}}, J.~A., {Johnstone}, D., {V{\'a}zquez-Semadeni}, E.,
  {Heitsch}, F., \& {G{\'o}mez}, G.~C. 2012, \apj, 756, 145

\bibitem[{{Seifried} \& {Walch}(2015)}]{Seifried:2015aa}
{Seifried}, D., \& {Walch}, S. 2015, \mnras, 452, 2410

\bibitem[{{Smith} {et~al.}(2014){Smith}, {Glover}, \& {Klessen}}]{Smith:2014aa}
{Smith}, R.~J., {Glover}, S.~C.~O., \& {Klessen}, R.~S. 2014, \mnras, 445, 2900

\bibitem[{{Stod{\'o}lkiewicz}(1963)}]{Stodolkiewicz:1963aa}
{Stod{\'o}lkiewicz}, J.~S. 1963, Acta Astron., 13, 30

\bibitem[{{Tafalla} \& {Hacar}(2015)}]{Tafalla:2015aa}
{Tafalla}, M., \& {Hacar}, A. 2015, \aap, 574, A104

\bibitem[{{Teyssier}(2002)}]{Teyssier:2002}
{Teyssier}, R. 2002, \aap, 385, 337

\bibitem[{{V{\'a}zquez-Semadeni} {et~al.}(2007){V{\'a}zquez-Semadeni},
  {G{\'o}mez}, {Jappsen}, {Ballesteros-Paredes}, {Gonz{\'a}lez}, \&
  {Klessen}}]{Vazquez-Semadeni:2007}
{V{\'a}zquez-Semadeni}, E., {G{\'o}mez}, G.~C., {Jappsen}, A.~K.,
  {Ballesteros-Paredes}, J., {Gonz{\'a}lez}, R.~F., \& {Klessen}, R.~S. 2007,
  \apj, 657, 870

\bibitem[{{Vishniac}(1994)}]{Vishniac:1994}
{Vishniac}, E.~T. 1994, \apj, 428, 186

\end{thebibliography}

\end{document}